\documentclass[fleqn,usenatbib]{mnras}

\usepackage{newtxtext,newtxmath}
\usepackage{pict2e} 
\usepackage{comment}
\usepackage[shortlabels]{enumitem}
\usepackage{hyperref}

\usepackage[T1]{fontenc}
\usepackage[capitalise]{cleveref}
\DeclareRobustCommand{\VAN}[3]{#2}
\let\VANthebibliography\thebibliography
\def\thebibliography{\DeclareRobustCommand{\VAN}[3]{##3}\VANthebibliography}

\usepackage{graphicx}	
\usepackage{amsmath}	
\usepackage{diagbox}
\usepackage{xcolor}

\title[Observing Optical kilonovae with medium size telescopes]{Optimising the observation of optical kilonovae with medium size telescopes\\}

\author[A. E. Camisasca et al.]{
A. E. Camisasca$^{1},$\thanks{E-mail: annaelisa.camisasca@unife.it}
I. A. Steele$^{2}$,
M. Bulla$^{1,3,4}$,
C. Guidorzi$^{1,3,5}$,
and
M. Shrestha$^{2,6}$
\\
$^{1}$Department of Physics and Earth Science, University of Ferrara, via Saragat 1, I--44122, Ferrara, Italy\\
$^{2}$Astrophysics Research Institute, Liverpool John Moores University, Liverpool Science Park IC2, 146 Brownlow Hill,
Liverpool L3 5RF, UK\\
$^{3}$ INFN -- Sezione di Ferrara, via Saragat 1, I--44122, Ferrara, Italy\\
$^{4}$ INAF -- Osservatorio Astronomico d’Abruzzo, via Mentore Maggini snc, 64100 Teramo, Italy\\
$^{5}$ INAF -- Osservatorio di Astrofisica e Scienza dello Spazio di Bologna, Via Piero Gobetti 101, I-40129 Bologna, Italy\\
$^{6}$Steward Observatory, University of Arizona, 933 North Cherry Avenue, Tucson, AZ 85721-0065, USA\\
}

\date{Accepted XXX. Received YYY; in original form ZZZ}

\pubyear{2022}

\begin{document}
\label{firstpage}
\pagerange{\pageref{firstpage}--\pageref{lastpage}}
\maketitle

\begin{abstract}
We consider the optimisation of the observing strategy (cadence, exposure time and filter choice) using medium size (2-m class) optical telescopes in the follow-up of kilonovae localised with arcminute accuracy to be able to distinguish among various kilonova models and viewing angles.
To develop an efficient observation plan, we made use of the synthetic light curves obtained with the Monte Carlo radiative transfer code \textsc{possis} for different kilonova models and as a function of different viewing angles and distances. By adding the appropriate photon counting noise to the synthetic light curves, we analysed four alternative sequences having the same total time exposure of 8 hours, with different time windows (0.5, 1, 2, 4 h), each with $i$, $r$, and $u$ filters, to determine the observing sequence that maximises the chance of a correct identification of the model parameters. We suggest to avoid $u$ filter and to avoid the use of colour curves.
We also found that, if the error on distance is $\leq 2\%$, $0.5$, 1, 2-hour time window sequences are equivalent, so we suggest to use 2-hour one, because it has 1 day cadence, so it can be easily realised. When the distance of the source is unknown, $0.5$~h time window sequence is preferable.

\end{abstract}





\section{Introduction}

Coalescences of neutron star binaries and black hole-neutron star systems lead to the formation of neutron-rich material. Such material undergoes rapid neutron capture nucleosynthesis (r-process) as it decompresses in space, leading to the creation of rare heavy elements such as gold and platinum \citep{Li98}. The radioactive decay of these unstable nuclei fuels a thermal transient known as "kilonova" (hereafter, KN; see \citealt{Metzger19_revKN} for a review).
On 2017 August 17, Advanced LIGO/Virgo made the first detection \citep{Abbott17} of gravitational waves (GW) from a binary neutron star merger, GW170817, simultaneously with the detection of  short gamma-ray burst (GRB) by Fermi \citep{Goldstein17} and INTEGRAL \citep{Savchenko17}: GRB 170817A. Eleven hours after the GW170817 trigger, an optical counterpart was discovered in the nearby (d = 40 Mpc) galaxy NGC 4993 \citep{Coulter17}.
The ultraviolet, optical, and near-infrared emission was consistent with being powered by the radioactive decay of nuclei synthesized in the merger ejecta by the r-process \citep{Villar17, Watson19,Domoto21,Kasliwal22}. This was the first time one source was detected both in GWs and electromagnetic (EM) radiation, and the first time spectroscopic evidence of a  KN was obtained \citep{Chornock17,Kasen17,Pian17b,Smartt17}.

 The study of a KN's rapid evolution can improve our understanding of the role of neutron star mergers in the origin of heavy elements. In addition, KN spectra encode key information to constrain the outflows that produced their electromagnetic emission. There has been only one confirmed case of KN detection in the form AT2017gfo and few other possible candidates such as KNe associated to GRB~130603B \citep{Tanvir17} and GRB~211211A \citep{Rastinejad22}. Hence, the whole community is working on various simulations to model the KN emission properties. There is a variety of predicted light curve features \citep[e.g.][]{Wollaeger18,Bulla19}. \citet{Klion21} and \citet{Nativi21} showed how the presence of a jet impacts the KN light curves and makes it brighter and bluer when viewed pole on. Thus, it is important to come up with efficient observational strategies to get the best observational data to constrain the properties from computational models.
 
In this work we aim to optimise the observing strategy for the optical followup of KNe to constrain the properties of the KN emission (viewing angle, mass of the different ejecta components and their velocities), once this has been identified and localised with arcminute accuracy, which enables observations with narrow field facilities. Arcminute accuracy can be achieved with current high energy instruments, such as the Burst Alert Telescope (BAT; \citealt{Barthelmy05}) on board the {\it Neil Gehrels Swift Observatory} \citep{Gehrels04}, or, in the near future, with {\it SVOM} \citep{SVOM22}, Einstein Probe  \citep{EinsteinProbeMission}, and in the next decade possibly {\it THESEUS} \citep{Amati21b}. Also, the advent of third-generation GW observatories, such as the Einstein Telescope (ET; \citealt{EinsteinTelescope}) and Cosmic Explorer (CE; \citealt{CosmicExplorer}), will lead to an accuracy in localisation better than 10~degrees$^{2}$ at $z < 3$, which is enough to enable prompt and efficient multiwavelength search for EM counterparts \citep{Ronchini22}. A GW detector capable of arcminute accuracy or better could be realised within the Voyage 2050 programme \citep{Baker21}.

We made extensive use of simulated multi-filter light curves (LCs) of KNe obtained with the POlarization Spectral Synthesis In Supernovae code (\textsc{possis}; \citealt{Bulla19}).  
Similar works recently carried out  \citep[e.g.][]{Scolnic18,Setzer19, Almualla21,Andreoni22, Chase22,Colombo22} focus on optimising strategies of wide field and follow up facilities to detect KNe. In the present work, instead, we aim to determine the optimal combinations of time exposure sequence and filters that help to constrain the model parameters with follow-up instruments.

We chose to study the specific case of small-medium class instruments; we considered 2 optical imaging cameras that are currently deployed at the 2-m fully robotic Liverpool Telescope \citep{Steele04}: MOPTOP \citep{Shrestha20} and IO:O \citep{Smith17}. We assume that a network of similar telescopes and instruments (e.g. \citealt{Tsapras09}) is located throughout a range of longitudes such that 24 hour coverage is available.  Given the interest in such sources, this assumption is reasonable in that most telescopes world-wide are likely to be involved in the followup of such rare events (e.g. \citealt{LCO13}).

In~\cref{Model parameters} we describe the characteristics of KN models generated with \textsc{possis}; in~\cref{Procedure with known distances} and~\cref{Results with known distances} we describe respectively the preliminary procedure and results obtained under the hypothesis of known source
distance; in~\cref{Procedure with distance uncertainty} and~\cref{sec:Results with distance uncertainty} we introduce the procedure and the results we adopted under the assumption of a distance uncertainty. We report our conclusions in~\cref{sec:Conclusion}.

\section{Model parameters}
\label{Model parameters}

\begin{figure}
    \includegraphics[width=\columnwidth]{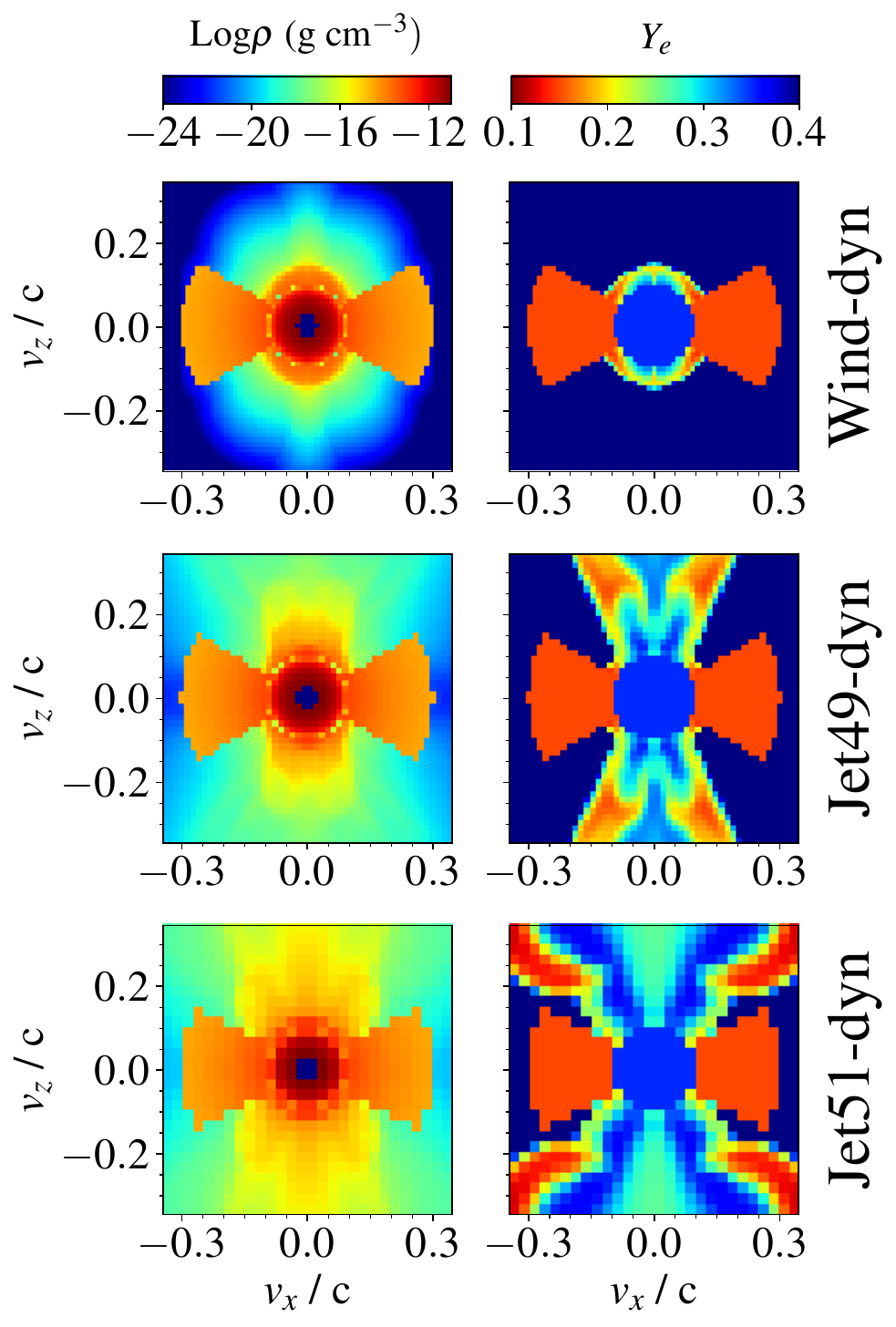}
\caption{Density (left) and $Y_e$ (right) distribution in the $x-z$ plane for the three models used in this study (\texttt{Wind-dyn}, \texttt{Jet49-dyn} and \texttt{Jet51-dyn} from top to bottom). Density maps are shown at $1$\,day after the merger.}
   \label{Models_density}
\end{figure}

We use KN models produced with \textsc{possis}, a 3-D Monte Carlo Radiative Transfer (MCRT) code that predicts photometric and polarimetric signatures of supernovae and KNe \citep{Bulla19}. The modelled ejecta are taken from \citet{Nativi21}, where a neutrino-driven wind as described in \citet{Perego14} was evolved assuming that either no jet (\texttt{Wind}), or a jet with a luminosity of $L_j = 10^{49}$~erg~s$^{-1}$ (\texttt{Jet49}), or a jet with $L_j = 10^{51}$~erg~s$^{-1}$ (\texttt{Jet51}) is launched. The wind mass is dominated by a secular component ejected 1\,s after the merger with $0.072\,M_\odot$. Unlike in \citet{Nativi21}, here we include an additional component to model dynamical ejecta. Specifically, we adopt an idealised geometry for this component, with a lanthanide-rich dynamical-ejecta component ($Y_e=0.15$ and velocities from 0.08 to 0.3c) from the grid in \citet{Dietrich20} and selecting the best-fit model to the KN of GW170817 (mass $0.005\,M_\odot$ and half-opening angle of $30^\circ$). 
These models are referred to as \texttt{Wind-dyn}, \texttt{Jet49-dyn}, and \texttt{Jet51-dyn} to distinguish them from those in \cite{Nativi21}. Figure~\ref{Models_density} shows density and $Y_e$ distributions for the three models. 

Radiative transfer simulations are carried out for the three models using the latest version of \textsc{possis} \citep{Bulla23}. Compared with the first version of the code \citep[][also used by \citealt{Nativi21}]{Bulla19}, the improved version assumes heating rates \citep{Rosswog22}, thermalization efficiencies \citep{Barnes16,Wollaeger18} and wavelength- and time-dependent opacities \citep{Tanaka20} that depend on the local properties of the ejecta as density, temperature and electron fraction. For each of the three models, we extract KN LCs for 11 different inclination angles for each model. Consequently, for a given distance and filter one has 33 different LCs. LCs are computed by \textsc{possis} from 0.1 to 30 days after the merger, but for this work we focus on the time windown from $1.0$ to $5.0$ days after the merger. We decided to ignore the code predictions earlier than 1 day after the merger because current opacity values assumed by \textsc{possis}  are likely affected by inaccuracies in the presence of highly ionised ejecta \citep{Tanaka20}. We do not consider LCs after 5 days due to the low value of flux.

The viewing angle $\theta$ is defined as the angle between the direction perpendicular to the merging plane and the line of sight. We used eleven values for the viewing angle separated by a constant step in cosine of $0.1$: $\cos{\theta}$ can assume the values 0, 0.1, 0.2, \ldots,1, with $\cos {\theta}=0$ corresponding to an observer in the merger plane (edge-on view) and $\cos {\theta}=1$ to an observer along the jet axis (face-on view).  We assumed the following range of values for distance: 20, 40, 80, 160, 250, and 350 Mpc. 

We chose to evaluate our results considering observations in the Sloan filter $i'$, $r'$, and $u'$ (hereafter referred to as $i$, $r$ and $u$).  These wavebands were chosen as being commonly available at most telescopes.  In particular we were keen to understand what (if any) additional value was added by carrying out $u$ band observations which are generally seen as more difficult than the $r$ and $i$ bands due to lower system throughputs and detector quantum efficiencies at near-ultraviolet wavelengths. 

\begin{figure*}
    \includegraphics[width=15 cm]{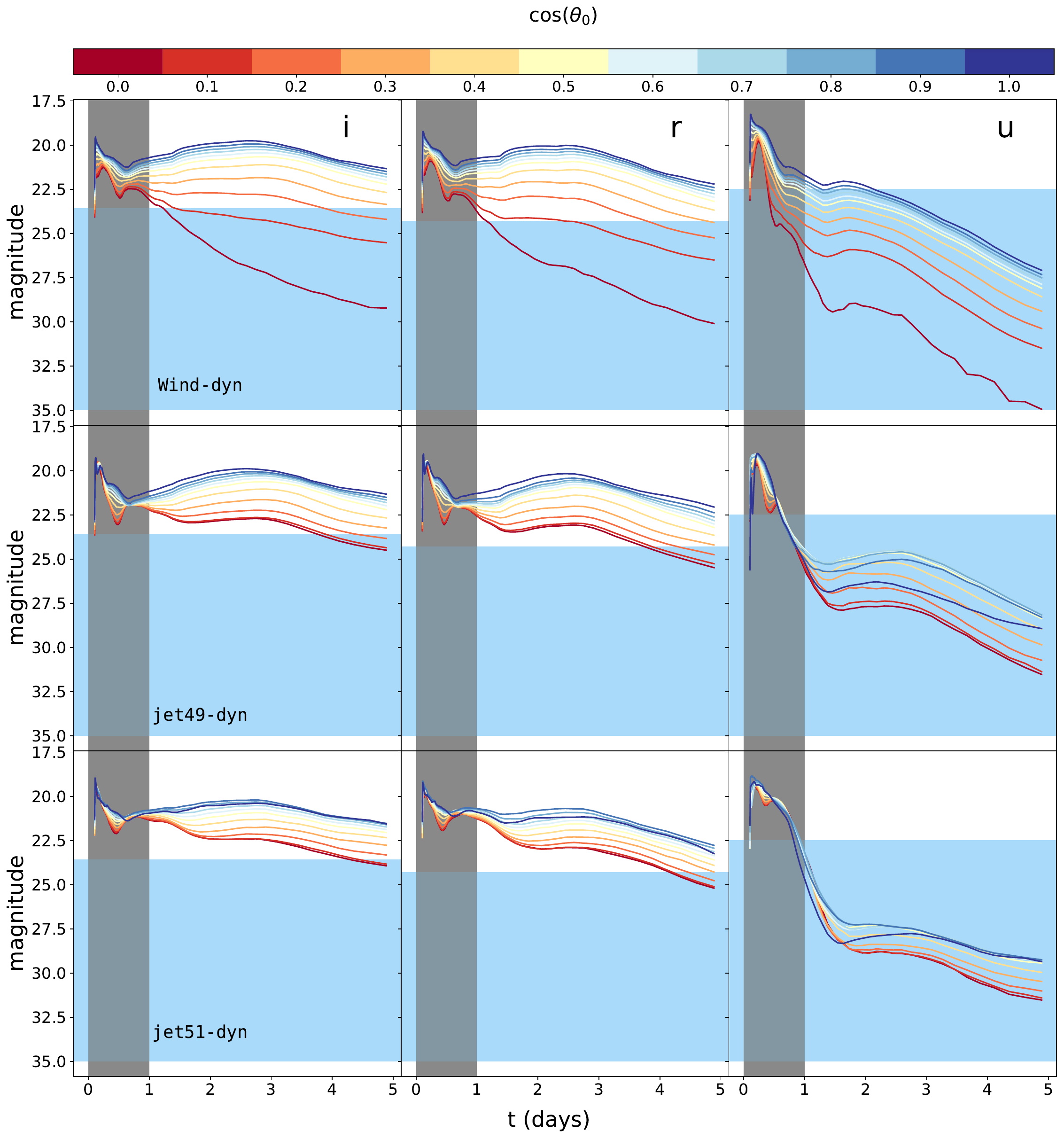}
\caption{LCs for different models and viewing angles; first column refers to i filter, second one to filter r, third to filter u; the source has a distance of 160 Mpc. Grey area corresponds to the first day after the merger: LCs are not considered due to the inaccuracy in estimating the opacity. The blue area corresponds to the values in magnitude higher than the limiting magnitude of each filter, obtained with a time exposure of 1 hour.}
   \label{Models_LCs}
\end{figure*}

In ~\cref{Models_LCs} we show the light curves obtained with d=160 Mpc for different filters, models, viewing angles. 
The KN brightness decreases going from the jet axis ($\cos \theta=1$) to the merger plane ($\cos \theta=0$) for all the three models, an effect that is caused by the presence of lanthanide-rich dynamical-ejecta material absorbing part of the escaping flux \citep[`lanthanide-curtain',][]{Kasen15,Wollaeger18}. The area highlighted in light blue in~\cref{Models_LCs} shows, for each filter, the magnitudes which are not detectable. Limiting magnitudes were obtained imposing a minimum signal to noise (SNR) threshold of 5. See~\cref{Appendix:A} for more details.

We point out that our study is restricted to a specific configuration in terms of ejecta properties (e.g. masses and compositions), since we do not aim to assess the ability of medium-size telescopes to constrain these properties, but rather to select the correct model and correct viewing angle. Extending this analysis to a large grid of models with different ejecta properties is beyond the scope of this paper and could be done in the future.

\section{Procedure with known distances}
\label{Procedure with known distances}
In order to find a reasonable time exposure sequence necessary to distinguish between different KN mo\-dels characterised by different viewing angles, we follow three main steps:
\begin{enumerate}
    \item we considered 4 different time exposure sequences, see ~\cref{sec:Time exposure sequences} for a detailed explanation.
    \item For a fixed distance and filter, we add the appropriate photon counting noise to the LC; the results depend on the time exposure sequence; we apply this step to all 33 LCs. Henceforward we will call the LC with noise "LCN"
    (see~\cref{Adding_noise_to_LC}).
    \item We compare LCN with the LCs without noise and we analyse how often we are able to identify the correct LC. This is done for each combination of  distance and filter. We repeat this for all the 33 different LCNs (see Section~\cref{Comparison_subsection}).
\end{enumerate}

In the first part the distance of the source is assumed to be known with negligible uncertainty, so we compare LCNs with LCs at the same distance.

\subsection{Time exposure sequences}
\label{sec:Time exposure sequences}
\begin{figure*}
    \includegraphics[width=15 cm]{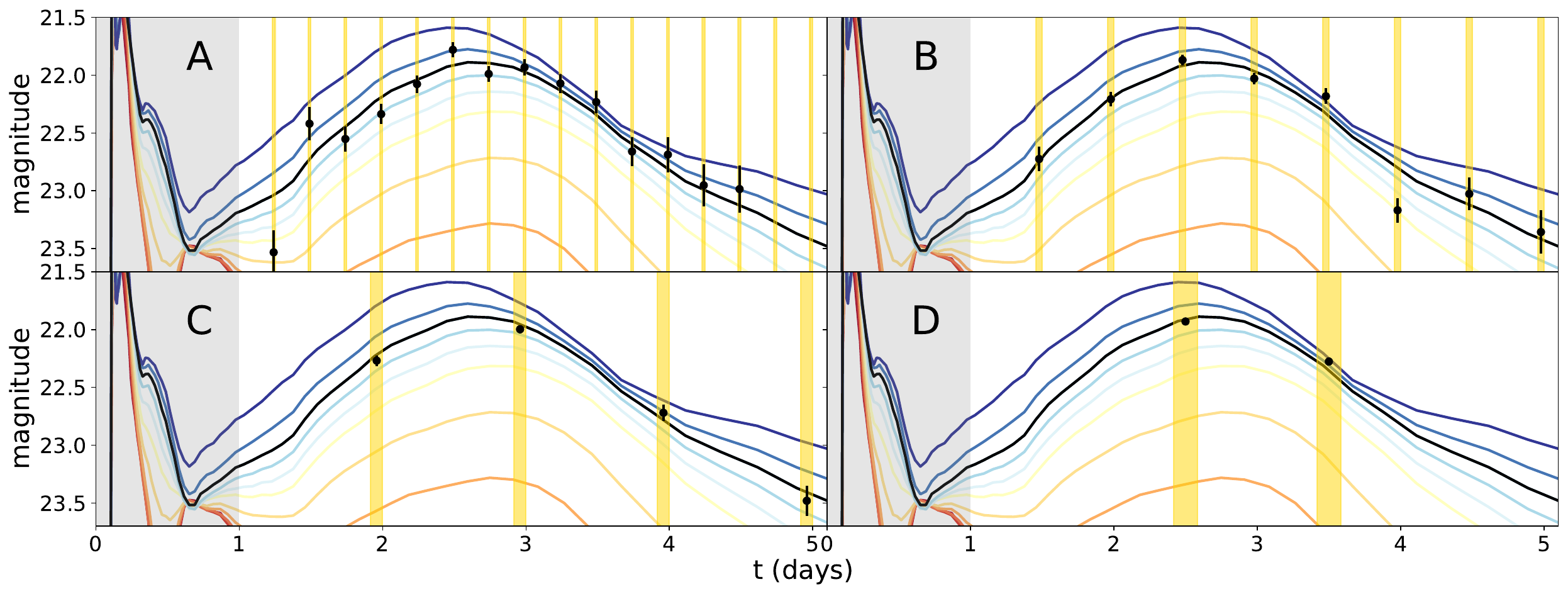}
    \caption{Each panel refer to a different time window sequence: A, B, C and D. In yellow, A, B, C and D time windows; in each plot, LCs referring to \texttt{jet49-dyn} model, d=350 Mpc, i filter. Black lines are LCs with $\cos \theta_{0}=0.8$, black points refer to the corresponding LCNs.}
   \label{fig:Time_windows}
\end{figure*}

We considered 4 different time exposure sequences (hereafter, referred to as A, B, C, D), each of them with a total net exposure of 8 hours.
\cref{tab:Time_exposure} reports the time windows and cadence for A, B, and C sequences. Sequence D requires a separate description: it consists of two 4-hour intervals 1 day apart. The exact times of the two observations are determined by maximising the difference between the two expected magnitude values taking into account the corresponding uncertainties. To this aim, for each instant we find the median of:
\begin{equation}
\frac{|m_{\rm model_{i}}-\overline{m}|}{\sigma_{m_{i}}}
\label{eq:median_m}
\end{equation}
where $m_{\rm model_{i}}$ is the magnitude of the $i$-th LC at that instant, $\overline{m}$ is the mean of all 33 LCs at that instant, $\sigma_{m_{i}}$ is 
\begin{equation*}
\label{eq:sigma_m_i}
    \sigma_{m_{i}}=\frac{1}{0.4  \sqrt{F_{i} \cdot t_{\rm exp}} \cdot \ln{(10)}}\;,
\end{equation*}
where $F_{i}$ is the photo-electron count expected in 1 s for the i-th model and $t_{\rm exp}=4$~h.
We sum the median (~\cref{eq:median_m}) obtained with different filters and distances and we find the mean value of this quantity in 4-hours intervals.We finally determine the maximum
of the sum of the value obtained in two 4-hours intervals 1 day
apart. In this way we obtain the intervals where the models are more different. Figure~\ref{fig:Time_windows} displays the resulting time windows.

\begin{table}
  \centering
    \begin{tabular}{ c| c c } 
     Name & duration  & cadence (d)\\
      & time exposure window (h) & \\
   \hline 
   A & 0.5 & 0.25 \\
   B & 1 & 0.5 \\
   C & 2 & 1 \\
   D & 4 & see~\cref{sec:Time exposure sequences} for a description\\     
 \end{tabular}
\caption{Duration of time exposure windows and cadence for 4 different time exposure sequences.}
\label{tab:Time_exposure}

\end{table}

\subsection{Adding noise to light curves}

\label{Adding_noise_to_LC}
We used A, B, C, D time exposure sequences to simulate  different light curves with noise for all combinations of filters, distances, viewing angles, and models. In more detail, at each time we calculated the expected photo-electron counts as $t_{\rm exp} (F+F_{\rm sky})$, where $F_{\rm sky}$ are the counts/s due to the sky (see \cref{Appendix:A} for more details). We then obtained the simulated counts $C_P$ by adding the statistical noise assuming the Poisson distribution.

The noise-affected flux of the $k$-th LC is calculated as follows:
\begin{equation}
F_{{\rm noise},k}=\frac{C_{P,k}}{t_{\rm exp}}-F_{\rm sky}
\label{F_noise}
\end{equation}
along with the corresponding magnitude:
\begin{equation}
m_{{\rm noise},k}\ =\ z_{P}-2.5\,\log_{10}{(F_{{\rm noise},k})}\;.
\label{m_noise}
\end{equation}
Equation \ref{m_noise} gives a generic LCN. Figure \ref{fig:Time_windows} shows the results of this step.

\subsection{Comparison between LC with noise and mo\-dels}
\label{Comparison_subsection}
We compare any given LCN with all of the 33 models and select the model which minimises the following $\chi^{2}$,
\begin{equation}
    \label{eq:Chi2}
    \chi^{2}(k,i)\ =\ \sum_{j=0}^{N_t} \left( \frac{m_{\rm model_{i}}(t_j)-m_{\rm noise_{k}}(t_j)}{\sigma_{m_{\rm noise_{k}}}(t_j)} \right)^{2} \cdot \frac{1}{N_t},
\end{equation}
where we are summing over the $N_t$ different data, $i$ and $k$ respectively identify LC and LCN, and $\sigma_{m_{\rm noise},k}$ is the uncertainty on $m_{{\rm noise},k}$, obtained by error propagating from $C_{P,k}$, using ~\cref{F_noise} and ~\cref{m_noise} and assuming $\sigma_{C_{P,k}}=\sqrt{C_{P,k}}$. It is:
\begin{equation}
\label{eq:sigma_m_noise}
    \sigma_{m_{{\rm noise},k}}\ =\ \frac{\sqrt{C_{P,k}}}{0.4  \cdot F_{{\rm noise},k} \cdot \ln{(10)} \cdot t_{\rm exp}}.
\end{equation}
In this way we obtain, for each model, viewing angle and distance (so, for each configuration), using 4 different time exposure sequences, the number of correct/incorrect matches.

\section{Results with known distances}
\label{Results with known distances}
\begin{figure*}

    \includegraphics[width=15 cm]{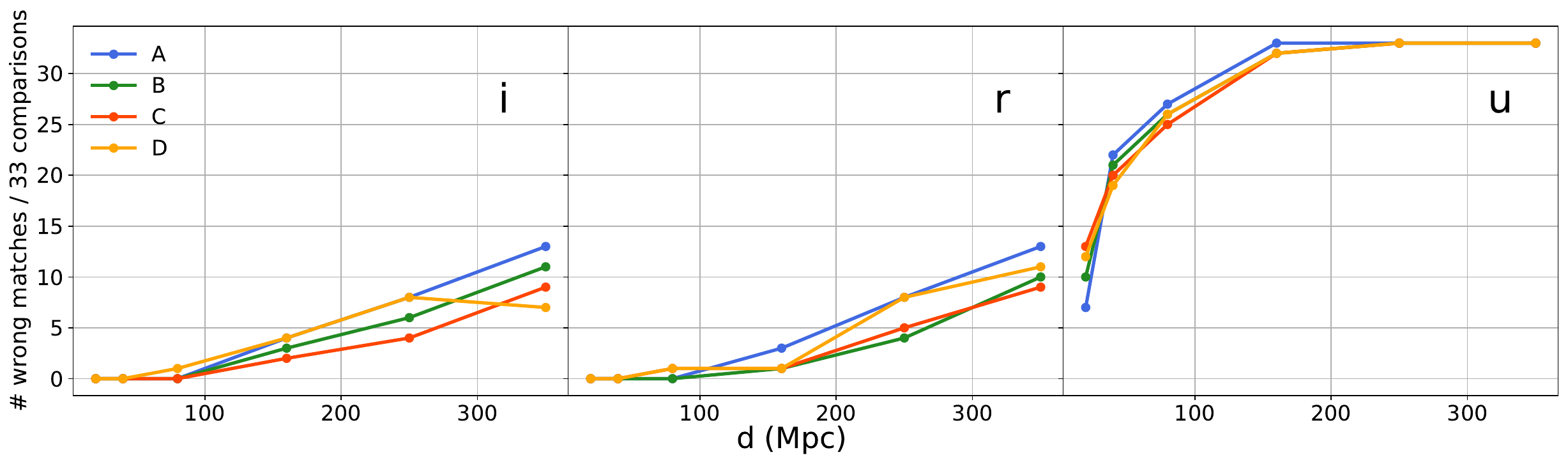}
\caption{Number of wrong matches as a function of distance for different filters and different time exposure sequences.}
\label{Comparison3procedures}
   \end{figure*}

\phantom{In this preliminary analysis, when we compare the light curve with noise with the 33 LCs models, we make the assumption that we know the source distance.}
Figure~\ref{Comparison3procedures} shows, for each distance and for each model, the number of incorrect matches out of 33 comparisons. Noticeably, it is better to avoid $u$ filter. Hereafter, in our analysis, we will consider only $i$ and $r$ filters.

We take note of three different kinds of mismatching errors between the simulated data points and the model lightcurve:
\begin{enumerate}[(I)]
    \item the most similar curve model corresponds to the simulated model, but the viewing angle is wrong;
    \item the most similar model turns out to be different from the original one;
    \item LCN is not detectable because the magnitude value is higher than the limiting value for each point of LCN.
\end{enumerate}

\begin{figure*}
\includegraphics[width=16 cm]{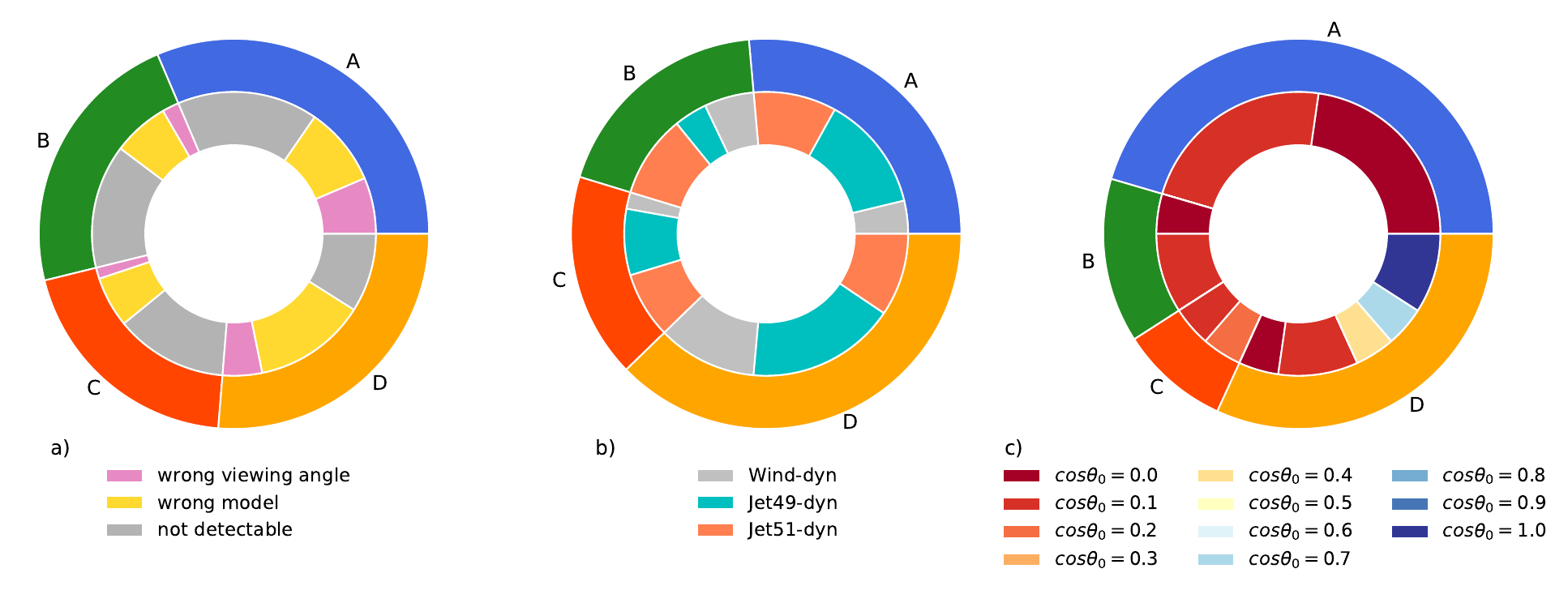}
\caption {For every different time exposure sequence, we analyse the type of mismatch that occurs in comparing the simulated curve with the models one. We considered both $i$ and $r$ filters. a)  In the outer ring, the number of wrong matches for each time exposure, in the inner ring the kind of mismatch. b) For every different time exposure sequence, we analyse which model is more difficult to detect. In the outer ring, the number of wrong matches for each time exposure, in the inner ring the number of mismatch for each kind of model. c) In the outer ring, the number of wrong matches for each time exposure, in the inner ring $\cos \theta_0$.}
\label{fig:First_pie}
   \end{figure*}

We summarise the results about the most common mismatches (i.e., mis-identifications) in the left pie of ~\cref{fig:First_pie}. The outermost ring corresponds to the different number of mismatches obtained with the 4 exposure combinations ($i$ and $r$ filters); the innermost one refers to the different kinds of mismatches. Overall, most of the mismatches are of type (III). The percentage of mismatches of type (II) is higher than that of type (I).
Type~(II) mismatches are shown in the central pie of~\cref{fig:First_pie} including both $i$ and $r$ filter; undetectable LCs are ignored. The number of mismatches is quite similar among all the kind of models and with every sequence. If we consider $i$ and $r$ filter individually, the results are similar; if we include the not detectable LCs, the number of mismatches with \texttt{Wind-dyn} model increases.
Let $\theta_{0}$ be the viewing angle of the LCN. We check if there is any particular value of $\theta_{0}$ for which we have most of the mismatches and how significant the mismatch is for the different viewing angle $\theta_{0}$. As shown in the right pie of ~\cref{fig:First_pie}, D time exposure sequence has a  wider range of starting angle that can bring to mismatches; with A, B and C most mismatches happen for $78^\circ\leq \theta_{0} \leq 90^\circ$. 
Analysing the difference between $\cos \theta_{0}$ and the value of $\cos \theta$ of the most similar LC, we find that with C time exposure sequence we always have $|\Delta \cos \theta|\leq 0.1$; with A and B we have more than 80\% of mismatches with $|\Delta \cos\theta|\leq 0.2$, with D it is 62\%.

The fact that the most of the mismatches refer to edge-on view and that $i$ and $r$ filter perform better than $u$ can easily be understood looking at~\cref{Models_LCs}: 
\begin{itemize}
\item [-] edge-on-view LCs are more significantly affected by statistical noise, since they have lower fluxes than face-on ones; also, they can partially or totally fade below the limiting magnitude to the point that they become undetectable;
\item [-]  $u$ filter LCs have lower fluxes and their limiting magnitude value is lower than $i$ and $r$ ones; these characteristics lead to a low performance of the $u$ filter. \end{itemize}

\section{Procedure with distance uncertainty}
\label{Procedure with distance uncertainty}
We now examine the realistic situation of non-negligible uncertainty on distance, and, in addition, the possibility that there is no information on the distance of the source as well as on the time of the merger. We restrict our analysis to $i$ and $r$ filters, due to the low performance of $u$. 

\subsection{1\% and 2\% error on distance}
We analysed the consequences of an error on distance of 1\% and 2\%. Such a level of accuracy in estimating the distance based on GW data alone appears to be feasible for a sizeable fraction of cases: third-generation gravitational-wave detector network will measure distances with an accuracy of $0.1\%$–$3\%$ for sources within $\leq 300$ Mpc (see Fig.~9 from \citealt{Gupta19}). 
To study this case, when we look for the matches between LCN and LCs, we shift the magnitude of the LC models due to the error on distance. We consider both a +1\% (+2\%) and -1\% (-2\%) error on distance. 
\subsection{Unknown distance, $\chi^{2}$-minimisation technique}
\label{Unknown distances}
Let us assume we have no information on distance as well as on the merger time. When we have to compare LCN with LCs, we start using the LCs-model with the intermediate distance of 160~Mpc (LC-160) and we shift LCs-160 both in time ($\Delta t$) and in magnitude ($\Delta m$), in order to find, among the 33 comparisons, $\Delta t$ and $\Delta m$ that minimise $\chi^{2}$ (~\cref{eq:Chi2}). 
Once we have $\Delta m$, we follow this procedure:
\begin{itemize}
    \item [-] we use $\Delta m$ to find an estimated distance ($d_{s}$) of the source;
    \item [-] we create a set of LC models with which to compare LCN, with a step of $0.1$ magnitudes between one model and the following one;
    \item [-] we choose the model with the nearest distance to $d_{s}$;
    \item [-] we compare LCN with the model at the most similar distance with $d_{s}$, shifting LCs both in time and magnitude to find the best match.
\end{itemize}
\subsection{Colour curves technique}
We adopted colour curves to try to limit the possible
effect of distance uncertainties. To create colour curve with noise (CCN), we add noise to LCs with different filters, then we subtract them.
Since CCNs have a dependence on distance (even if small), when we compare CCNs with colour curve models (CCs), we compare them with CCs-model at the intermediate distance of 160~Mpc (CC-160).  

\section{Results with distance uncertainty}
\label{sec:Results with distance uncertainty}
\subsection{Single filter technique}
\begin{figure*}
    \includegraphics[width=15 cm]{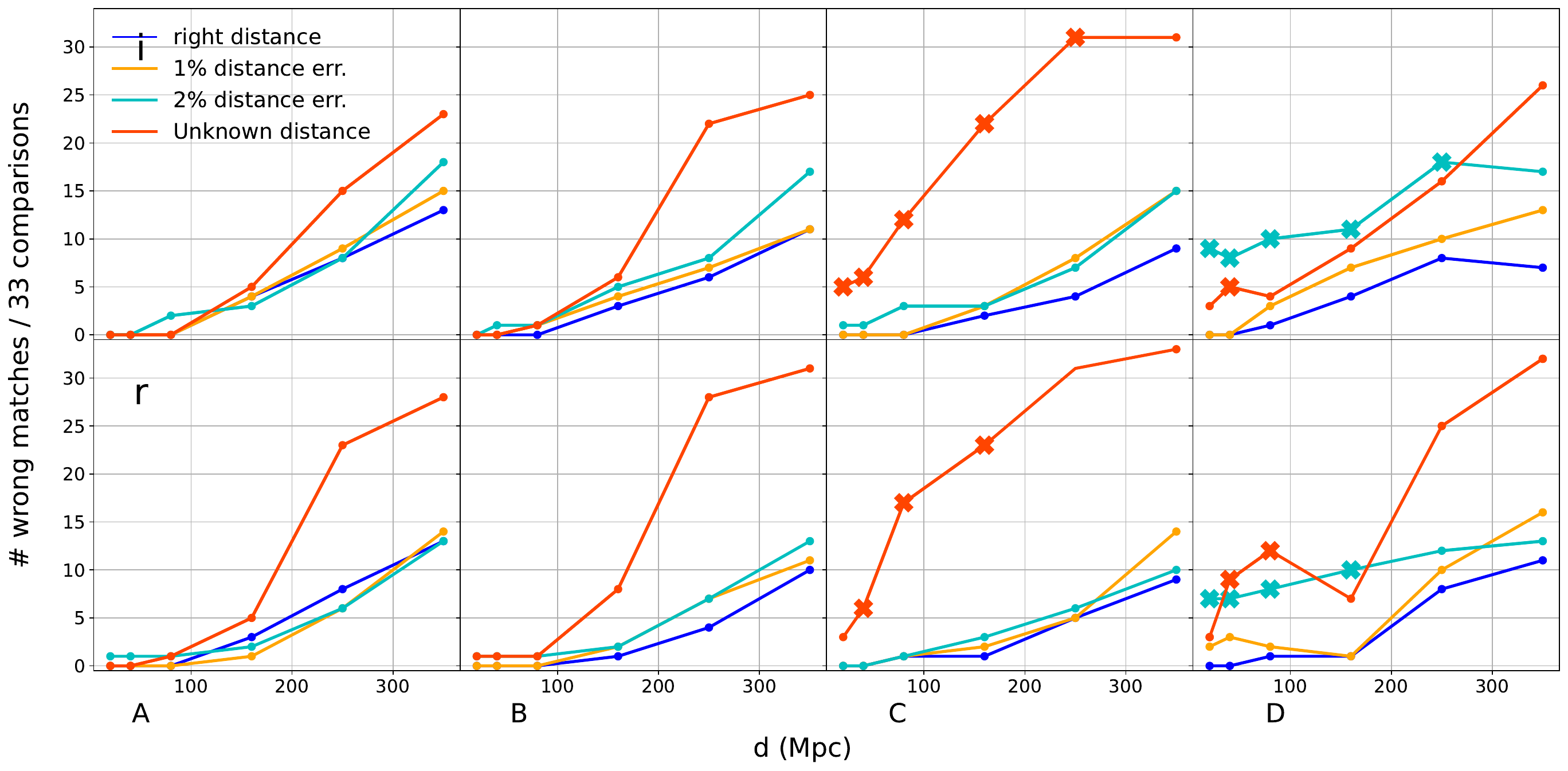}
\caption{The number of wrong matches that occurs in 33 comparisons with A, B, C, and D time exposure sequences with a known distance (blue line), with an error of 1\% on distance (yellow line), 2 \% (cyan line) and without any information about the distance and the time of the merger (orange line). Points marked with ``x'' refer to values that are significantly different from the best value obtained with other time sequences.}
\label{fig:Comparison_error_on_distance}
   \end{figure*}

In Figure~\ref{fig:Comparison_error_on_distance} we present the number of wrong matches that occur in 33 comparisons with A, B, C, and D time exposure sequences, using filters individually. 
We reported the results obtained without error on distance and we compare it with what we obtain with an error of 1\%, 2\%, and without information on distance and time of the merger. For 1\% (and 2\%) error on distance we plot the highest number of mismatches between +1\% and -1\% (+2\% and -2\%).

For each filter, each distance, we compare the results obtained with A, B, C, and D checking if the number of mismatches within each time exposure sequence is compatible with the best results obtained within the limits of Poisson statistics.\footnote{When we compare the results of two observations $N_1$ and $N_2$, we assume that the numbers of wrong matches are independently Poisson distributed. Consequently, |${N_2-N_1}$| is the absolute value of a Skellam-distributed random variate. We calculate the probability of having $\ge|N_2-N_1|$ assuming as expected value for the common Poisson distribution the mean value of $N_1$ and $N_2$.
When the probability is $<5$\% the two numbers are considered significantly different.} 
In Figure~\ref{fig:Comparison_error_on_distance} we marked with ``x'' the cases that are significantly different from the best value obtained with other time sequences; if the error on distance is $\leq \%2$, no particular statistical differences emerge between sequences A, B, and C. When the distance is unknown, C and D time windows should be avoided.
Hereafter, we restrict our analysis to A and C time exposure sequences, since C, with 1 day cadence, can easily be carried out with a single telescope, but should be avoided when no information on the source distance is available.
The comparison between $i$ and $r$ shows that they are equivalent. 

\subsubsection{Error on distance $\leq 2\%$}
\begin{figure*}
    \includegraphics[width=16cm]{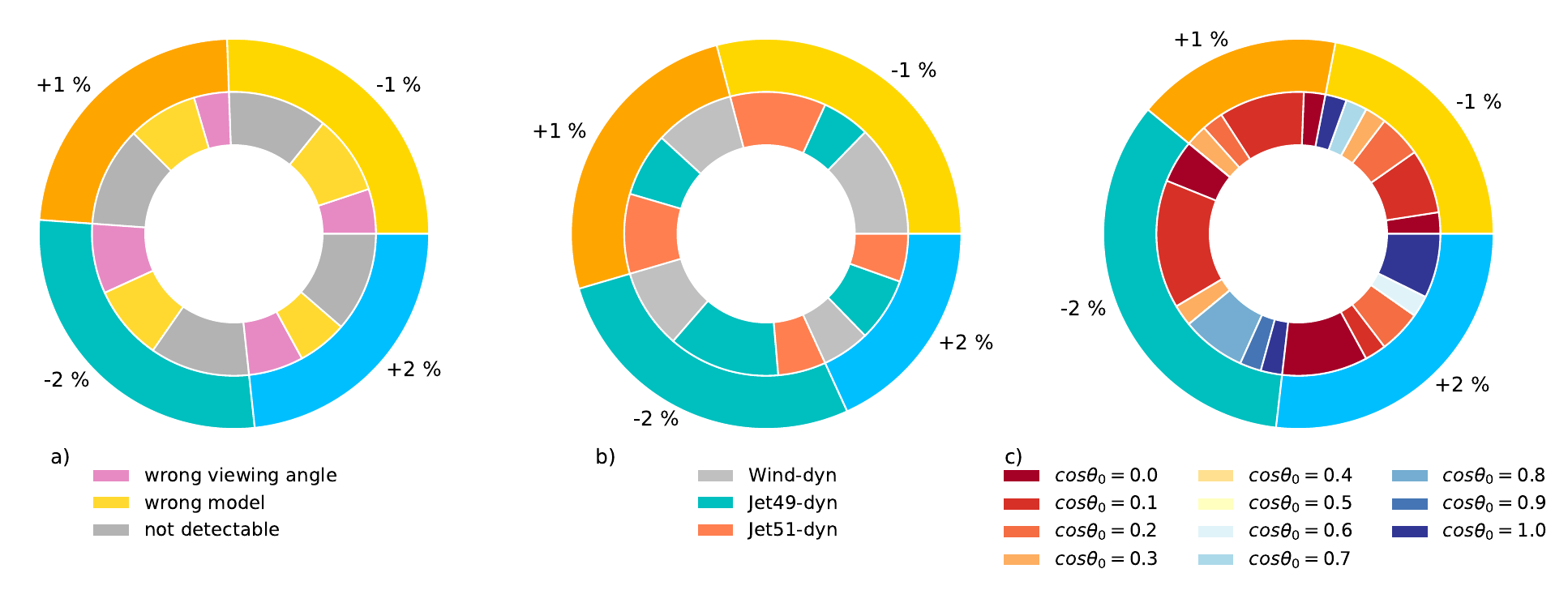}
\caption{Using time sequence C, we consider the number of wrong matches we have when we make an error on distance of $\pm$1\%, $\pm$2\%. a. In the outer ring, the number of wrong matches for each error on distance, in the inner ring the kind of mismatch. b. For every different model of the simulated source, we analyse how many mismatches we have. c. For every different observational direction, we analyse the number of mismatches.}
\label{fig:Pie_error_on_distanceC}
   \end{figure*}

Focusing on time sequence C with error on distance  $\leq 2\%$, Figure~\ref{fig:Pie_error_on_distanceC} shows that the number of mismatches of type (I), (II), (III) is mostly the same.
The number of mismatches concerning the models are equally distributed between \texttt{Wind-dyn}, \texttt{Jet49-dyn}, and \texttt{Jet51-dyn}, provided that the light curve can be detected. If we consider also not detectable LCNs, \texttt{Wind-dyn} model mismatches increase (i.e. \texttt{Wind-dyn} model would be harder to be detected).
With an error on distance of 1\%, the majority of the mismatches are in the interval $73^\circ\leq \theta_{0} \leq 90^\circ$; this interval becomes wider for increasing errors on distance.
When the error on distance is $\leq 2 \%$, both with $i$ and $r$ filter we have $|\Delta \cos{\theta}|\leq 0.2$ for all the mismatches.

\subsubsection{Unknown distance}
If we do not have information about the distance, in order to have the lowest number of mismatches, it is recommendable to use A time window sequence; with this sequence the number of mismatches of type (I), (II), and (III) is similar; also the mismatches concerning the models are equally distributed among \texttt{Wind-dyn}, \texttt{Jet49-dyn}, and \texttt{Jet51-dyn}. Regarding viewing angle mismatches, they occur with the same frequency for every $\theta_{0}$; furthermore, $|\Delta \cos{\theta}|\leq 0.2$ for 82\% of mismatches. 

\subsubsection{Focus on viewing angle estimation}
For each combination of distance and of its error we adopted the following procedure: for each viewing angle $\theta_{0}$ we determined the uncertainty on the estimated viewing angle $\theta_{\rm est}$, using either C or A time sequence respectively for error on distance $\leq2\%$ and for unknown distance. 
Then we took the largest uncertainty among all the values of $\theta_0$: in this way, we associated to any combination of distance and error on it with a conservative uncertainty in the estimated viewing angle, as the result of any possible value of $\theta_0$.
\cref{tab:st_deviation} reports the results.

If the error on distance is $\leq 2\%$, the error on $\theta_{\rm est}$ is always $\leq 7^\circ$; these errors implicitly assume that the inaccuracies intrinsic to the \textsc{possis} models and its assumptions are negligible. In practice, should this be no more the case, an independent estimate of the viewing angle, combined with the errors reported in \cref{tab:st_deviation}, could help to constrain the \textsc{possis} accuracy, thus providing useful feedback to tweak and refine the code itself.

\begin{table}
  \centering
    \begin{tabular}{ c| c c c} 
   d (Mpc)   & 1\% distance err. & 2\% distance err. & Unknown distance\\
   \hline 
   20 & $3^\circ$ & $4^\circ$ & $11^\circ$ \\
   40 & $2^\circ$ & $7^\circ$ & $7^\circ$ \\
   80 & $1^\circ$ & $4^\circ$ & $5^\circ$ \\
   160 & $3^\circ$ & $4^\circ$ & $12^\circ$ \\
   250 & $2^\circ$ & $4^\circ$ & $14^\circ$ \\
   350 & $7^\circ$ & $4^\circ$ & $25^\circ$ \\    
 \end{tabular}
\caption{The higher standard deviation on $\theta_{est}$ for different distances.}
\label{tab:st_deviation}

\end{table}

\subsection{Colour curve technique}
\begin{figure}
\centering
    \includegraphics[width=7cm]{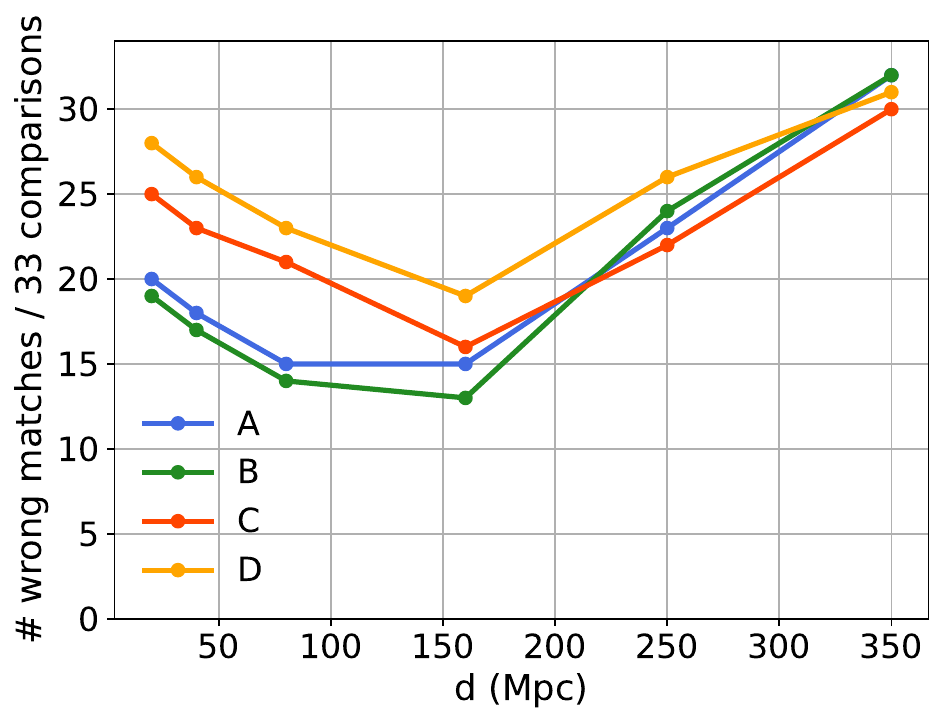}
\caption{Number of wrong matches as a function of distance for different time exposure sequence using colour curves $i$-$r$.}
\label{Comparison_colors}
   \end{figure}

For each time windows sequence, we analysed the results obtained comparing CCN with CC-160 for $i$-$r$ CCs. As we can see in~\cref{Comparison_colors}, this procedure gives a higher number of wrong matches than single filter techniques; this is due to the fact that the use of 2 LCs increases the possibility that, in a given instant, there is at least one undetectable LC. Moreover, uncertainties on both curves combine and lower the SNR; furthermore, LCs in i and r are really similar and there is really little viewing angle dependence in i-r colour.
We do not consider i-u and r-u CCs due to u filter outcomes; using another filter combined with i and r might lead to better results.
Finally, since there is a slight dependence of CCs on distance, we use in the comparison CC-160; this makes the match more difficult when the distance is highly different from 160 Mpc.

\section{Conclusions}
\label{sec:Conclusion}
The aim of the paper was finding the best strategy to characterise accurately localised KNe with follow-up small-medium size optical telescopes. We found that the use of the $u$ filter should be avoided, due to the high number of mismatches with all the time window sequences considered in this work (see~\cref{Comparison3procedures}). Even a procedure with $i$-$r$ colour curve is not as convenient as one might think: it gives a higher number of wrong matches than single filter technique, due to the fact that the use of 2 LCs increases the possibility that, in a moment, there is at least one undetectable LC.
Alternative time window sequences sharing the same total net exposure and with at least 4 observations and a maximum cadence of 1~day, are essentially equivalent, provided that  the error on distance is $\leq 2\%$. Consequently, we suggest to use one day cadence sequence, because it can be easily realised.
If the distance of the source is unknown, short cadence ($\le 0.5$ day) sequences are preferable. 

Finally, we demonstrated that, for any distance considered in the present analysis (from 20 to 350~Mpc) and an error on distance $\leq 2\%$, the viewing angle is estimated very accurately: the correct value is always compatible with the estimated one within uncertainties, with an error that is always $\leq 7^\circ$. This means that an independent measurement of the viewing angle could help to constrain the accuracy of \textsc{possis}, providing useful information to refine the code itself. In addition, more stringent constraints on the viewing angle can better reduce the distance-inclination angle degeneracy in GW data, and, consequently, lead to a more accurate estimate of the distance and of the Hubble constant $H_{0}$ (e.g., \citealt{Guidorzi17,Dhawan20}, see \citealt{Bulla22b} for a review). 

\section*{Acknowledgments}
We are grateful to the Referee for their careful reading and constructive comments
which helped us to improve the paper.
A.E.C. thanks I.A.S. for the hospitality at the Astrophysics Research Institute of Liverpool JM University and the significant research opportunity and acknowledges the University of Ferrara for financial support under the programme "Call for mobility scholarship for periods at European and extra European Institutions". The radiative transfer simulations with \textsc{possis} were performed on resources provided by the Swedish National Infrastructure for Computing (SNIC) at Kebnekaise.

\section*{Data Availability}

The $uri$ light curves for the six models used in this study will be made available at \url{https://github.com/mbulla/kilonova_models}. The \textsc{possis} code used to simulate the light curves is not publicly available.




\bibliographystyle{mnras}
\bibliography{alles_grbs} 




\appendix

\section{Exposure time formula for a single filter}
\label{Appendix:A}

We calculate $F_{\rm lim}$, defined as the minimum photo-electron count collected in 1~s to have a detectable signal, assuming a limiting signal to noise SNR$_{\rm lim}=5$, through the following equation:
\begin{equation*}
 {\rm SNR}_{\rm lim}= \frac{F_{\rm lim}\ t_{\rm exp}}{\sqrt{F_{\rm lim} t_{\rm exp}+F_{\rm sky} t_{\rm exp}}}   
\end{equation*}
with 
\begin{equation*}
 F_{sky}= 10^{0.4(z_{P}-m_{\rm sky})}\ A
\end{equation*} 
where $z_{P}$ is the instrument zero point referred to a particular filter (the magnitude corresponding to one detected photo-electron per second), $m_{\rm sky}$ is the sky magnitude in 1 arcsec$^{2}$ and $A$ is the area of the photometric aperture used. We used $z_{P}$ and $m_{\rm sky}$ values as suggested at the Liverpool Telescope website,\footnote{\url{https://github.com/LivTel/ETC_calcs/blob/master/NRT\_calc.html}} assuming $m_{\rm sky}$ as intermediate between a dark and a bright sky (\cref{zP}). Since the typical La Palma seeing is 0.75 arcsec, we adopted an aperture diameter two times that value (i.e. 1.5 arcsec), which yields $A=1.8$ arcsec$^{2}$. 

\begin{table}
  \centering
    \begin{tabular}{ c| c c} 
filter & $z_{p}$ & $m_{\rm sky}$ (mpsas) \\
\hline
$i$ & 25.06 & 17.3 \\
$r$ & 15.39 & 18.4 \\
$u$ & 21.00 & 18.0 \\
 \end{tabular}
\caption{$z_{P}$ and $m_{\rm sky}$ values}
\label{zP}

\end{table}



\bsp	
\label{lastpage}
\end{document}